\documentclass[twocolumn,tighten]{aastex62}
\usepackage{textcomp}
\usepackage{gensymb}

\usepackage[hang,raggedright,tight]{subfigure}
\usepackage{longtable}
\usepackage{tabularx}
\usepackage[figuresright]{rotating}
\usepackage{float}
\usepackage[toc,page]{appendix}

\usepackage{gensymb}
\usepackage{threeparttable}
\usepackage{diagbox}
\usepackage{pbox}
\usepackage{booktabs}
\usepackage{multirow}  
\usepackage{makecell}
\usepackage{natbib}
\usepackage{amsmath,bm}
\usepackage{CJKutf8}

\shorttitle{AP test}
\shortauthors{Dong et al.}

\begin{document}
\title{Galaxy Clustering Topology in Observations and Cosmological Simulations}

\author[0000-0003-0296-0841]{Fuyu Dong}
\altaffiliation{dfy@ynu.edu.cn}
\affiliation{South-Western Institute for Astronomy Research, Yunnan University, Kunming, Yunnan 650500, People's Republic of China}
\affiliation{Yunnan Key Laboratory of Survey Science, Yunnan University, Kunming, Yunnan 650500, People's Republic of China}
\affiliation{Shanghai Key Laboratory for Particle Physics and Cosmology, 200240, People’s Republic of China}
\author[0000-0001-9521-6397]{Changbom Park}
\altaffiliation{cbp@kias.re.kr}
\affiliation{School of Physics, Korea Institute for Advanced Study (KIAS), 85 Hoegiro, Dongdaemun-gu, Seoul, 02455, Republic of Korea}

\begin{abstract} 
We present a genus topology analysis of the large-scale distribution of galaxies in the Sloan Digital Sky Survey Data  (SDSS), comparing the observed topology with mock galaxy samples from the Horizon Run 4 (HR4) $N$-body simulation and four cosmological hydrodynamical simulations. We construct nine volume-limited samples and measure their genus curves at Gaussian smoothing scales of $4$--$12~h^{-1}{\rm Mpc}$. Overall, the modern galaxy models reproduce the observed genus statistics substantially better than the HGC, HOD, and SAM models examined by Choi et al. (2010). Among the models considered here, the HR4 Most-Bound-Halo Particle galaxy-assignment model shows somewhat better overall agreement with the SDSS topology than IllustrisTNG300 and FLAMINGO. The observed scale dependences of the genus shift parameter $\Delta\nu_f$ and void abundance parameter $A_V$ are somewhat steeper than those predicted by the simulations, but their joint discrepancy from HR4 is not statistically significant. The smaller-volume TNG100 and EAGLE realizations show significant excesses in $A_C$, at $4.5\sigma$ and $7.0\sigma$, respectively, relative to the observational uncertainties. When the expected cosmic variance of the simulations is taken into account, they do not necessarily indicate failures of the galaxy-formation prescriptions, but they caution that individual small-volume realizations may not be representative of the observed galaxy topology. Overall, modern galaxy formation models reproduce the observed topology substantially better than earlier models, while residual differences remain that can be tested more precisely with denser future surveys.
\end{abstract}

\section{Introduction}
Modern galaxy formation models are calibrated against observed galaxy
properties, including luminosity and stellar mass functions and two-point
clustering statistics. However, agreement with these statistics does not
uniquely determine the spatial distribution of galaxies: models with similar
two-point clustering may differ substantially in the higher-order arrangement
of galaxies within the large-scale structures in the universe (LSS).
Higher-order statistics therefore provide important additional tests of
galaxy formation models.

The topology of the galaxy distribution provides such a complementary probe.
The genus statistic \citep{1986ApJ...306..341G,1987ApJ...319....1G}
quantifies the connectivity of overdense and underdense regions and is
sensitive to phase correlations not captured by two-point statistics.
On large scales, topology preserves information on the primordial density
field and tests Gaussian random-phase initial conditions
\citep{1986ApJ...304...15B}, while on smaller scales it is modified by
gravitational evolution, galaxy bias, and galaxy formation processes.

The topology of galaxy clustering has been extensively studied in redshift
surveys
\citep{1989ApJ...340..625G,2008ApJ...675...16G,
1992ApJ...392L..51P,1992ApJ...387....1P,2001ApJ...553...33P,
2005ApJ...633...11P,1992MNRAS.256..477M,1994ApJ...420..525V,
1994ApJ...421....1R,1997ApJ...489..457P,1998MNRAS.297..777C,
2002ApJ...570...44H,2002ApJ...580..663H,2002PASJ...54..707H,
2003PASJ...55..911H,2007MNRAS.375..128J,2009MNRAS.394..454J},
revealing systematic departures from Gaussian topology due to nonlinear
evolution and biased galaxy formation. \citet{2010BKAS...35Q..82C}
tested galaxy formation models using the SDSS DR7 main galaxy sample by
comparing the observed genus statistics with mock catalogs based on the halo-galaxy one-to-one correspondence (HGC), halo occupation distribution (HOD), and semi-analytic (SAM) models. Among these, the HGC model was significantly more successful than the HOD and SAM models in reproducing the observed topology, but none reproduced all of its features simultaneously. Significant discrepancies remained in the genus amplitude and shift and in the abundances of isolated clusters and voids, demonstrating the discriminating power of topology among galaxy formation models.

Since that study, cosmological simulations and galaxy modeling have advanced substantially. The Most-Bound-Halo Particle (MBP) galaxy assignment method
\citep{2016ApJ...823..103H} applied to Horizon Run 4 (HR4) $N$-body simulation accurately reproduces a broad range of observed galaxy clustering statistics, while modern cosmological hydrodynamical simulations form galaxies self-consistently. Large-volume simulations such as IllustrisTNG300 and FLAMINGO \citep{2017MNRAS.465.3291W,2018MNRAS.473.4077P,2023MNRAS.526.4978S}, and the widely used IllustrisTNG100 and EAGLE simulations \citep{2015MNRAS.446..521S,2015MNRAS.450.1937C}, reproduce many observed galaxy properties, but their galaxy topology has not been systematically tested against observations.

These advances raise a natural question: do modern galaxy formation models reproduce the observed topology of galaxy clustering as well as they reproduce conventional galaxy statistics? Because topology probes higher-order spatial information, discrepancies may reveal aspects of galaxy bias, environmental processes, satellite evolution, or baryonic feedback that are difficult to identify from one- and two-point statistics alone.

In this paper, we revisit the topology test using the same SDSS DR7 Main Galaxy sample analyzed by \citet{2010BKAS...35Q..82C}. We compare the observed galaxy topology with that predicted by the MBP galaxy model applied to the HR4 $N$-body simulation and by galaxies formed in the IllustrisTNG300 and FLAMINGO cosmological hydrodynamical simulations. We also examine IllustrisTNG100 and EAGLE, whose smaller volumes make their topology more susceptible to cosmic variance. By comparing the genus curve, its amplitude and shift, and the abundances of isolated clusters and voids, we test whether contemporary galaxy formation models reproduce the higher-order topology of the observed galaxy distribution and characterize any remaining discrepancies.

\section{The genus statistic}
\label{sec:genus}
We measure the topology of the smoothed galaxy density field using the genus statistic \citep{2005ApJ...633...11P}. For an isodensity contour surface $S$, the genus is related to the integrated Gaussian curvature through the Gauss–Bonnet theorem,
\begin{equation}
    G = N_\text{holes} - N_\text{isolated}=-\frac{1}{4\pi}\int_S\kappa da,
\end{equation}
where $\kappa$ is the local Gaussian curvature, $da$ denotes the area element on the surface $S$.
We use the genus density $g(\nu)=G(\nu)/V$, where $V$ is the sample volume. 

For a Gaussian random field, the genus density has the analytic form
\begin{equation}
\label{gG}
g_G(\nu)=g(1-\nu^2)e^{-\nu^2/2},
\end{equation}
where $\nu\equiv(\rho-\bar{\rho})/\sigma$ is the density threshold, with
$\sigma=\langle(\rho-\bar{\rho})^2\rangle^{1/2}$. The amplitude of genus density is
\begin{equation}
g=\frac{1}{(2\pi)^2}
\left(\frac{\langle k^2\rangle}{3}\right)^{3/2},
\end{equation}
where
\begin{equation}
\label{eq:k2}
\langle k^2\rangle=
\frac{\int P(k)W^2(k)k^2\,d^3k}
{\int P(k)W^2(k)\,d^3k},
\end{equation}
and $W(k)$ is the Fourier transform of the smoothing kernel, taken to be
Gaussian in this analysis.

When analyzing the samples, rather than parameterizing the contour by the density threshold itself, we use the volume-fraction threshold \(\nu_f\), defined such that the volume fraction above a given density equals that of a Gaussian random field at the corresponding threshold. This allows us to probe non-Gaussianity beyond that contained in the one-point density distribution.

We characterize each measured genus curve using four genus-related statistics \citep{2005ApJ...633...11P}: the genus amplitude \(g\), the shift parameter \(\Delta\nu\), and the void- and cluster-related parameters \(A_V\) and \(A_C\). The amplitude $g$ is obtained by fitting the Gaussian genus curve to the measured genus curve over $-1\le\nu\le1$. The remaining parameters quantify deviations from this reference Gaussian curve:
\begin{equation}
\Delta\nu=\frac{\int_{-1}^1g(\nu)\nu d\nu}
{\int_{-1}^{1}g_G(\nu) d\nu},
\end{equation}
\begin{equation}
A_V=\frac{\int_{-2.2}^{-1.2}g(\nu) d\nu}
{\int_{-2.2}^{-1.2}g_G(\nu) d\nu},
\end{equation}
\begin{equation}
A_C=\frac{\int_{1.2}^{2.2}g(\nu) d\nu}
{\int_{1.2}^{2.2}g_G(\nu) d\nu}.
\end{equation}
Here, $g_G$ denotes the same best-fitting Gaussian genus curve defined above. $A_V$ and $A_C$ characterize the relative abundances of isolated voids and clusters, respectively. For a Gaussian random field, \(\Delta\nu=0\) and \(A_V=A_C=1\). A negative \(\Delta\nu\) corresponds to a meatball, or cluster-dominated, shift, whereas a positive value indicates a bubble-like shift.

To compare samples with different smoothing scales, we use the dimensionless genus amplitude $gR_G^3$. For a Gaussian field, this quantity is determined by the shape of the power spectrum. 

\section{Observational and simulation data}
\label{sec:data}

\subsection{SDSS DR7 galaxy sample}
\label{sec:genus}

\begin{table}
\centering
\caption{
Volume-limited samples of SDSS galaxies.
\label{Table-sample}}
\centering
\begin{tabularx}{0.4\textwidth}{c|c|c|c|c}
\hline
\hline
\multirow{2}*{Sample}&\multirow{2}*{$M_{r,\rm cut}$} &\multirow{2}*{$z_{\rm max}$} &\multirow{2}*{${\bar d}$} &\multirow{2}*{${N_{\rm mock}^{\rm HR4}}$}\\ 
    &        &         &{}[$h^{-1}\rm{Mpc}$]{} &\\
\hline
G1  & -18.89 & $\,0.065\,$ & 4  &215\\
G2  & -19.77 & $\,0.095\,$ & 5  &125\\
G3  & -20.19 & $\,0.115\,$ & 6  &125\\
G4  & -20.51 & $\,0.131\,$ & 7  &64 \\
G5  & -20.72 & $\,0.145\,$ & 8  &64 \\
G6  & -20.893& $\,0.154\,$ & 9  &27 \\
G7  & -21.03 & $\,0.165\,$ & 10 &27 \\
G8  & -21.16 & $\,0.174\,$ & 11 &27 \\
G9  & -21.26 & $\,0.182\,$ & 12 &27 \\       
\hline
\end{tabularx}
\begin{tablenotes}
\small
\item Note. The redshift range of each sample is $0.025<z<z_{\rm max}$. $M_{r,\rm cut}$ is the absolute $r$-band magnitude, and ${\bar d}$ is the mean galaxy separation.
\end{tablenotes}
\end{table}

\begin{figure}
    \centering
     \subfigure{
     \includegraphics[width=1\linewidth, clip]{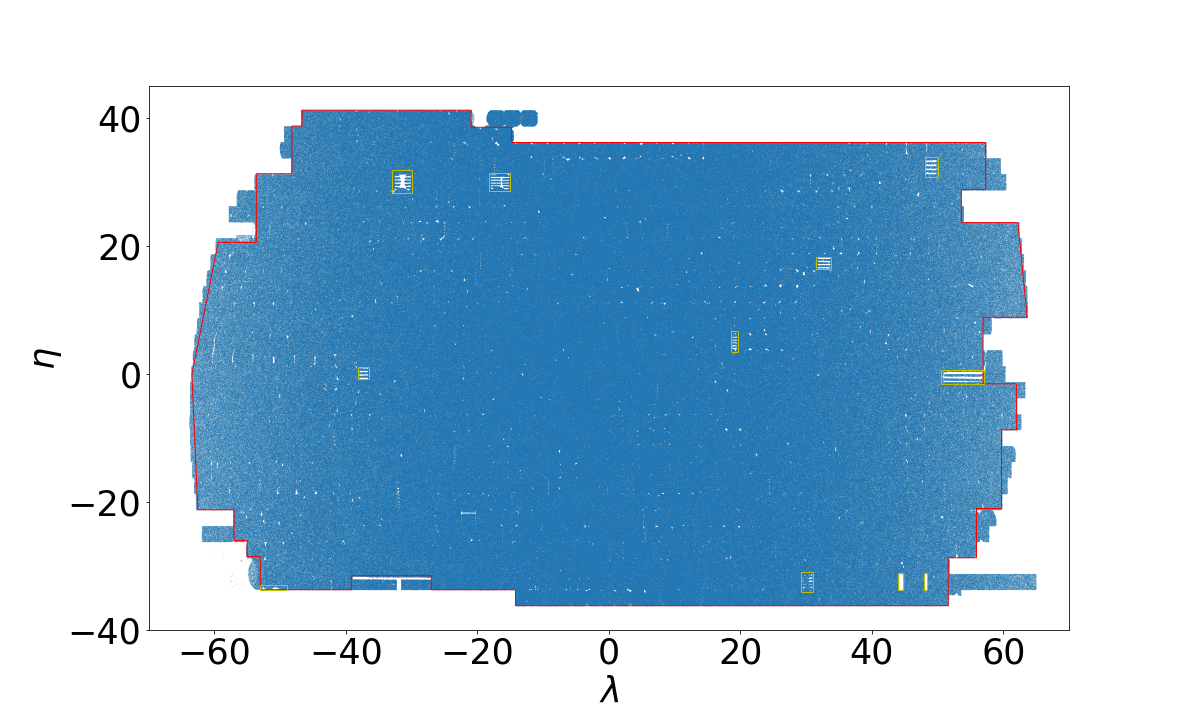}}
   \caption{Angular definition of the SDSS sample in the survey coordinate plane ($\lambda, \eta$), presented using random points generated with the SDSS DR7 angular selection function. The red solid lines define the boundaries of the analysis regions, while the yellow solid lines mark the inner regions that are excluded owing to their significant incompleteness in the angular selection function. 
   } 
    \label{fig:DR7}
\end{figure}

We use the Korea Institute for Advanced Study Value-Added Galaxy Catalog of \citet{2010BKAS...35Q..82C}, constructed from the SDSS DR7 Main Galaxy Sample by supplementing the NYU-VAGC with additional redshifts from the literature to improve spectroscopic completeness, particularly in regions of high galaxy surface density. The catalog provides spectroscopic redshifts, photometric properties, and an angular selection function accounting for spatially varying completeness after inclusion of the additional redshifts. For the present analysis, we restrict the angular coverage as shown in Figure~\ref{fig:DR7} to reduce the surface-to-volume ratio, excluding regions with significant incompleteness and narrow, protruding survey boundaries. 

We then construct nine volume-limited samples using progressively brighter \(r\)-band absolute-magnitude thresholds over \(0.025<z<z_{\rm max}\) (see Table~1). The thresholds are chosen such that the mean galaxy separation ${\bar d}$ increases from \(4\,h^{-1}{\rm Mpc}\) for G1 to \(12\,h^{-1}{\rm Mpc}\) for G9.

\subsection{Horizon Run 4 simulation}
Several physical and observational effects can cause the measured genus curve to deviate from the Gaussian form, including nonlinear structure formation, redshift-space distortions (RSDs), survey selection and masking, and Poisson sampling noise. We use the HR4 mock catalogs to quantify observational and sampling effects and to validate our genus estimator, and then compare the topology of HR4 and the hydrodynamical simulations with that of the SDSS galaxy samples.

The HR4 simulation has a box size of $3150\,h^{-1}{\rm Mpc}$ and contains $6300^3$ dark matter particles \citep{2015JKAS...48..213K}. It adopts the WMAP 5-year cosmology, with $(\Omega_{\Lambda},\Omega_m,\Omega_b,h,\sigma_8,n_s) =(0.74,0.26,0.044,0.72,0.794,0.96)$, and simulation details are given in Table.\ref{Table-hydro}. Galaxies are assigned to the most-bound dark matter particles (MBPs) of physically self-bound halos/subhalos \citep{2016ApJ...823..103H}, with satellite survival determined using the modified merger-timescale model of \citet{Jiang_2008}.

For comparison with the observations, galaxy positions are transformed from real to redshift space by including the line-of-sight peculiar velocities. We use the snapshot at $z=0.1$ and construct mock catalogs matching the geometries of the SDSS samples; the number of mocks for each sample is given in Table~\ref{Table-sample}.
 
\subsection{Hydrodynamical simulations}
\begin{table*}
\centering
\caption{
Main characteristics of the cosmological simulations used for the genus analysis.
\label{Table-hydro}}
\centering
\begin{tabular}{c|c|c|c|c|c|c}
\hline
\hline
&HR4 &TNG100-1   &TNG300-1  &TNG300-DMO  &FLAMINGO L1\_m8 &EAGLE L0100 \\
\hline
$\Omega_mh$                           &0.1872                &0.2092                 &0.2092                 &0.2092            &0.2084                &0.2081\\ 
Simulation code                       &GOTPM                 &AREPO                  & AREPO                 &AREPO             & SWIFT                &GADGET-3 \\
Simulation type                       &gravity               &gravity+hydro          &gravity+hydro          &gravity           &gravity+hydro         &gravity+hydro \\
Numerical method                      &TreePM                &Moving mesh            &Moving mesh            &TreePM            &\textsc{sphenix sph}  &\textsc{anarchy sph}\\
Box size [cMpc]                       &$4375$                &$110.7$                &$302.6$                &$302.6$           &$1000$                & $100$ \\   
$\lambda_{\rm Nyq,IC}$[cMpc]          &$1.389$               &$0.122$                &$0.242$                &$0.242$           &$0.556$               &$0.133$\\
$M_{\rm DM}[M_\odot]$                 & $1.25\times10^{10}$        &$\sim7.5\times10^6$    &$\sim5.9\times10^7$    &$7\times10^7$     &$\sim7.06\times10^8$  & $\sim9.70\times10^6$ \\
$M_{\rm gas}[M_\odot]$                &-                     & $\sim1.4\times10^6$   &$\sim1.1\times10^7$    &-                 & $\sim1.34\times10^8$ & $\sim1.81\times10^6$ \\
\multirow{2}*{Grav. resolution}        &\multirow{2}*{69.4 ckpc}  &\multirow{2}*{1.48 ckpc}    &\multirow{2}*{2.95 ckpc}  &\multirow{2}*{2.95 ckpc}  &\multirow{2}*{11.2 ckpc}   &\multirow{2}*{2.66 ckpc}\\
                                      &{}{}        &{}0.74 pkpc{}         &{}1.48 pkpc{}           &{}1.48 pkpc{}     &{} 2.85 pkpc {}       &{}0.70 pkpc{}\\ 
\multirow{2}*{Galaxy identification}  &\multirow{2}*{\textsc{psb+mbp}}          &\multirow{2}*{\textsc{subfind}}      &\multirow{2}*{\textsc{subfind}}       &\multirow{2}*{\textsc{subfind}}  &\multirow{2}*{\textsc{hbt-herons}}    &\multirow{2}*{\textsc{subfind}}\\
                                      &                      &                      &                       &{}+\textsc{l-galaxies}{}&+\textsc{soap}    & \\
\multirow{2}*{Galaxy mass threshold}  &\multirow{2}*{$11.47 ~(M_{\rm halo})$}&\multirow{2}*{$9.79 ~(M_*)$}&\multirow{2}*{$9.55~(M_*)$}&\multirow{2}*{$9.67~(M_*)$}&\multirow{2}*{$11.24~(M_*)$}&\multirow{2}*{$9.62~(M_*)$}\\
{}log[$M/(h^{-1}M_\odot)]${} & & & & & &\\
$\bar{d}\;[h^{-1}{\rm Mpc}]$          &$\sim4$               &$\sim4$               &$\sim4$                &$\sim4$           &$\sim9$               &$\sim4$   \\
\hline
\end{tabular}
\begin{tablenotes}
\small 
\item Note. 
$\lambda_{\rm Nyq,IC}$ denotes the shortest wavelength represented by
independent modes in the initial density field and is twice the initial
particle/grid spacing. It is distinct from the particle-mass resolution
and gravitational softening length.
For simulations in which the gravitational softening changes from comoving
to proper coordinates, the two entries give the comoving softening length
and the maximum proper softening length, respectively.
MBP denotes the most-bound halo particle; SUBFIND and HBT-HERONS identify
gravitationally bound substructures, while SOAP measures the properties of
the HBT-HERONS structures.
$\bar{d}$ denotes the mean galaxy separation of the densest
sample for the genus analysis.
\end{tablenotes}
\end{table*}

We use four cosmological hydrodynamical simulations to test whether galaxy populations formed with baryonic physics reproduce the observed topology: TNG100-1 and TNG300-1 from IllustrisTNG \citep{2017MNRAS.465.3291W,2018MNRAS.473.4077P}, 
FLAMINGO L1\_m8 \citep{2023MNRAS.526.4978S}, and EAGLE Ref-L0100N1504 \citep{2015MNRAS.446..521S,2015MNRAS.450.1937C}. 
Their main properties are summarized in Table~\ref{Table-hydro}. All adopt flat $\Lambda$CDM cosmologies but differ in cosmological parameters, volume, resolution, hydrodynamical method, and galaxy-formation prescriptions.

IllustrisTNG uses the AREPO moving-mesh method, FLAMINGO uses SWIFT with the SPHENIX SPH formulation, and EAGLE uses a modified version of GADGET-3 with ANARCHY SPH. Their galaxy-formation models include radiative cooling, star formation, stellar evolution and chemical enrichment, stellar feedback, black-hole growth, and AGN feedback, with additional physics such as magnetic fields in TNG and massive neutrinos in FLAMINGO. These unresolved processes are modeled with subgrid prescriptions, and differences in numerical methods, feedback models, and calibration can produce different galaxy populations and galaxy--halo connections, potentially affecting the predicted genus statistics.

Galaxy identification also differs among the simulations. HR4 identifies gravitationally self-bound (PSB) structures and uses the MBP method to assign central and satellite galaxies based on their merger trees, while IllustrisTNG and EAGLE identify substructures with SUBFIND \citep{10.1046/j.1365-8711.2001.04912.x,10.1111/j.1365-2966.2009.15034.x}. In FLAMINGO, bound substructures are identified with HBT-HERONS \citep{2018MNRAS.474..604H,2025JOSS...10.8252M}, and their properties are measured with SOAP.\footnote{FLAMINGO provides galaxy and subhalo properties using several definitions and apertures. In this work, we use the stellar properties associated with bound substructures identified by HBT-HERONS and measured with SOAP.}

The simulation volumes limit the smoothing scales over which statistically meaningful comparisons can be made. TNG100-1 and EAGLE are used only at $R_G=4\,h^{-1}{\rm Mpc}$, while TNG300-1 is used at $R_G\leq7\,h^{-1}{\rm Mpc}$ and the larger lower-resolution simulation FLAMINGO L1\_m8 at $R_G\geq9\,h^{-1}{\rm Mpc}$. 
\section{Results}
\label{sec:result}
In this section, we present the genus curves for the SDSS galaxy samples over a series of Gaussian smoothing lengths, characterize their scale dependence using the genus-density amplitude \(g\), shift parameter \(\Delta\nu_f\), cluster abundance parameter \(A_C\), and void abundance parameter \(A_V\), and compare these measurements with those from mock SDSS samples constructed from cosmological simulations.

We examine the RSD and survey-mask effects in Appendix~\ref{sec:validade}, galaxy assignment and baryonic effects through comparisons with simulations in this section, and Poisson sampling effects in Sec.~\ref{sec:noise}.

\subsection{Comparison Between SDSS Observation and HR4 Simulation}
\begin{figure*}
    \centering
     \subfigure{
     \includegraphics[width=1\linewidth, clip]{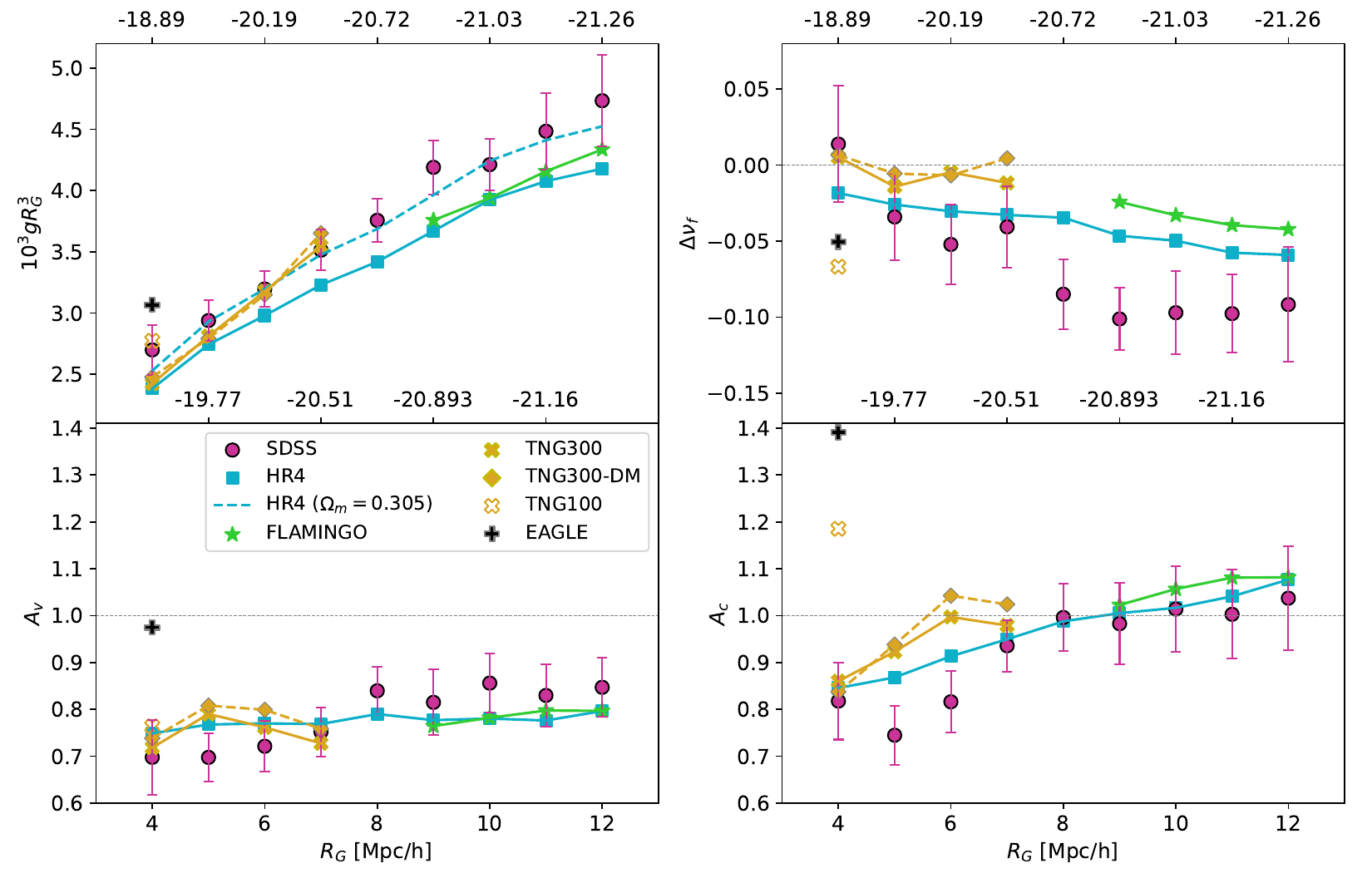}}
   \caption{Genus-related amplitude and shape parameters measured from nine volume-limited samples. Measurements for SDSS galaxies are shown as magenta circles, while those for HR4 mock galaxies are shown as green squares. For comparison, genus curves from TNG300 (including both hydrodynamical and dark-matter-only simulations) and FLAMINGO are also included, distinguished by yellow saltires, yellow diamonds, and green stars, respectively. We also present TNG100 and EAGLE at $R_G=4h^{-1}{\rm Mpc}$ for comparison, denoted as open yellow saltires and black crosses, respectively. The green dashed line in the first panel shows the scaled HR4 genus amplitude by changing $\Omega_m$ from 0.26 to 0.305, which is the best-fit value from the SDSS eAP test \citep{2026ApJ...998...66D}. 
   } 
    \label{fig:shape}
\end{figure*}

\begin{figure}
    \centering
     \subfigure{
     \includegraphics[width=1\linewidth, clip]{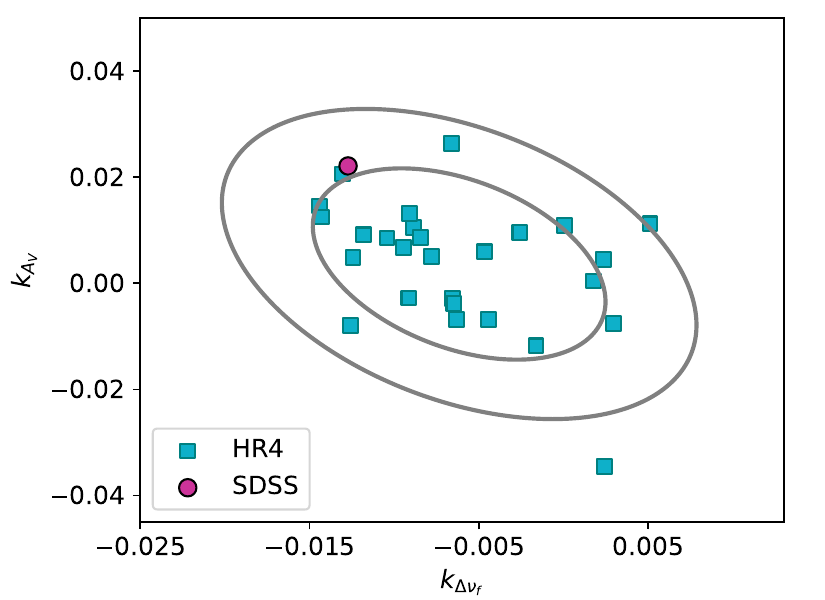}}
   \caption{The distribution of the $\Delta\nu_f$ slope compared with that of $A_V$, where the genus is measured from the SDSS (magenta point) and HR4 mocks (blue squares) while the smoothing scale is varied from 4 to 12 $h^{-1}{\rm Mpc}$.} 
    \label{fig:slope-pdf}
\end{figure}

Figure~\ref{fig:shape} presents the genus-related statistics as a function of the Gaussian smoothing scale \(R_G\) for the observational samples and simulations. The results for the nine volume-limited SDSS samples listed in Table~1 are shown in pink, with error bars estimated from the corresponding HR4 mock samples. The absolute-magnitude limit of each sample is indicated on the \(x\)-axis. To maximize the statistical power, the smoothing length is chosen to be nearly equal to the mean galaxy separation of each sample, and the pixel size used to construct the galaxy number-density field is set to less than \(R_G/4\) \citep{2014ApJS..212...22K}.

Mock SDSS samples are constructed from HR4 by applying the survey mask and angular selection function and matching the observed galaxy number density in redshift space through an appropriate halo-mass cut. The number of mocks for each sample is given in Table~1. The mean genus-related statistics of the HR4 mock samples are shown as blue squares.

The observed genus amplitude \(gR_G^3\) in the SDSS samples increases monotonically with \(R_G\). The shift parameter \(\Delta\nu_f\) becomes progressively more negative toward the brighter and sparser samples, while both \(A_V\) and \(A_C\) increase with \(R_G\).
These systematic trends reflect the combined effects of scale-dependent clustering, luminosity bias, and shot noise.

Comparing the observational results with HR4, we note the following:
\begin{itemize}
\item
The genus amplitude \(gR_G^3\) is systematically higher in the observations than in HR4, although the difference is only at the level of about \(1-2\sigma\). This discrepancy may indicate that the matter power spectrum adopted in HR4 is too steep over the relevant scales. When the HR4 prediction is rescaled to a cosmology with \(\Omega_m=0.305\) or $\Omega_m h=0.220$, as recently measured by \cite{2026ApJ...998...66D}, instead of the HR4 value of 0.26, the genus amplitude agrees very well with the observations (blue dashed line).

\item For $\Delta\nu_f$, SDSS measurements are consistent with HR4 at
$R_G\le7\,h^{-1}{\rm Mpc}$, but become more negative than the HR4
predictions at larger smoothing scales.

\item Both $A_V$ and $A_C$ are statistically consistent with the HR4 predictions over the comparison range, although the SDSS $A_V$ shows some scale dependence that is not seen in HR4.
\end{itemize}

Overall, the comparison with HR4 shows substantially better agreement than that obtained with the HGC galaxy-assignment schemes examined by \citet{2010BKAS...35Q..82C}.

This improvement suggests that the MBP galaxy-assignment prescription provides a more realistic representation of galaxy populations in \(N\)-body simulations.

To assess whether the discrepancies between SDSS and HR4 noted above for \(\Delta\nu_f\) and \(A_V\) are statistically significant, we perform a test using the HR4 mock samples. We construct 27 sets of HR4 mocks that reproduce the overlapping spatial geometries of the G1--G9 samples, thereby preserving the correlations among the samples. For each set, we fit the \(R_G\) dependence of  \(A_V\) and \(\Delta\nu_f\) with linear relations and measure their slopes, \(k_{A_V}\) and \(k_{\Delta\nu_f}\), respectively. Figure~\ref{fig:slope-pdf} shows $(k_{A_V}, k_{\Delta\nu_f})$ of the 27 mock sets and SDSS (magenta circle).
We model the two-dimensional distribution of  \((k_{A_V}, k_{\Delta\nu_f})\) with a bivariate Gaussian, and calculate the probability of obtaining a slope pair at least as discrepant from the HR4 mean as the observed SDSS pair (see the \(1\sigma\) and \(2\sigma\) contours in Fig.~\ref{fig:slope-pdf}). The resulting probability is \(7.3\%\). Thus, the simultaneously more negative \(\Delta\nu_f-R_G\) slope and more positive \(A_V-R_G\) slope observed in SDSS are marginally consistent with statistical fluctuations due to cosmic variance and shot noise in the HR4 mocks, when the correlations among the G1--G9 samples are taken into account.

\subsection{Comparisons with hydro-simulations}
\label{sec:hydro-4mpc}
\begin{figure}
    \centering
     \subfigure{
     \includegraphics[width=1\linewidth, clip]{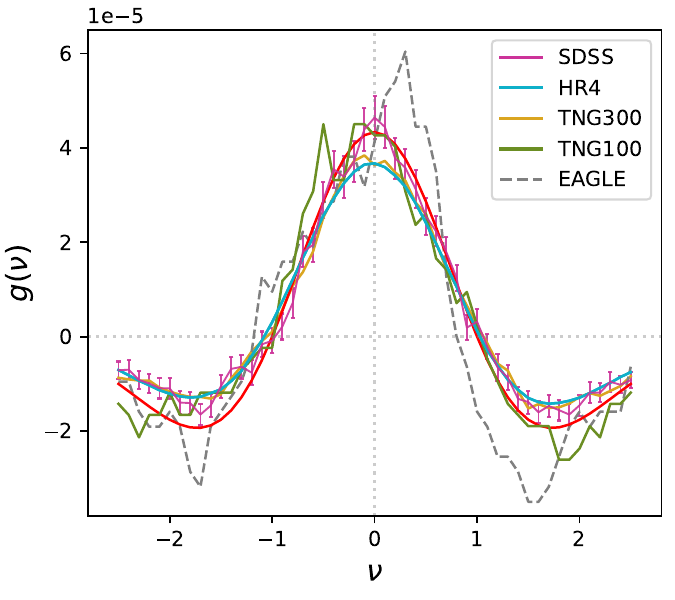}}
   \caption{Genus curve measured for G1 sample ($M_r < -18.89$) with $R_G=4h^{-1}{\rm Mpc}$. In addition to SDSS DR7, HR4 and TNG300, we show the measurements from TNG100 and EAGLE simulations for comparison. 
   }
    \label{fig:genus-G1}
\end{figure}

\begin{table}
\centering
\caption{
Genus-related amplitude and shape parameters measured for the G1 sample from observation and simulations, with $R_G=4h^{-1}Mpc$.
\label{Table-G1}}
\centering
\begin{tabular}{ccccc}
\hline
\hline
simulation & $10^3gR_G^3$ & $\Delta\nu$ & $A_V$ & $A_C$\\
\hline

\multirow{2}*{SDSS}   &\multirow{2}*{2.700} &\multirow{2}*{0.014}&\multirow{2}*{0.698} &\multirow{2}*{0.818} \\
                     &{}$\pm0.204${}       &{}$\pm0.038${}       &{}$\pm0.080${}       &{}$\pm0.082${} \\ 
HR4 &2.385 &-0.018 &0.749 &0.846 \\        
\multirow{2}*{TNG300}&\multirow{2}*{2.425} &\multirow{2}*{0.005} &\multirow{2}*{0.719} &\multirow{2}*{0.860} \\
                        &{}$(\pm0.092)${}    &{}$(\pm0.029)${}       &{}$(\pm0.043)${}     &{}$(\pm0.043)${} \\
\multirow{2}*{TNG100}&\multirow{2}*{2.775} &\multirow{2}*{-0.067}&\multirow{2}*{0.764} &\multirow{2}*{1.185} \\
                     &{}$(\pm0.587)${}       &{}$(\pm0.095)${}       &{}$(\pm0.200)${}       &{}$(\pm0.216)${} \\
\multirow{2}*{EAGLE} &\multirow{2}*{3.067} &\multirow{2}*{-0.050}&\multirow{2}*{0.975} &\multirow{2}*{1.391} \\
                     &{}$(\pm0.680)${}       &{}$(\pm0.110)${}       &{}$(\pm0.232)${}       &{}$(\pm0.250)${}\\
\hline
\end{tabular}
\begin{tablenotes}
\small 
\item Note. The error bar for TNG300 is estimated via the jackknife method by iteratively excluding one-eighth of the simulation volume. The error bar for TNG100 is estimated by dividing the TNG300 volume into 27 independent subvoluems, while the error bar for EAGLE is estimated by scaling the TNG100 statistical error by the square root of the ratio of their simulation volumes.
\end{tablenotes}
\end{table}

Figure~\ref{fig:shape} also includes measurements from the hydrodynamical simulations. Such simulations employ empirical subgrid prescriptions for unresolved baryonic processes, whose parameters are calibrated against a limited set of key observations, such as the cosmic star-formation-rate density and the galaxy stellar mass function. The simulated galaxy populations are therefore constrained to reproduce the observables used for calibration, while agreement with independent observables provides a more stringent test of the galaxy-formation model. The topology of the galaxy distribution offers such a test.

Both TNG300 and FLAMINGO yield \(A_V\) values very close to those of HR4. Larger discrepancies are seen, however, in \(\Delta\nu_f\) and \(A_C\), for which the hydrodynamical simulations generally lie farther from the observations than HR4. At \(R_G=5\,h^{-1}{\rm Mpc}\), the TNG300 value of \(A_C\) is higher than the observed value by about \(2.8\sigma\), while at \(R_G=9\,h^{-1}{\rm Mpc}\), the FLAMINGO value of \(\Delta\nu_f\) is more positive than the observation by about \(3.7\sigma\). Overall, the topologies predicted by TNG300 and FLAMINGO deviate somewhat more from the observations than does HR4, with some of the discrepancies reaching marginal statistical significance.

Because the volumes of TNG100 and EAGLE are much smaller than those of the other simulations, we compare them with SDSS and the other simulations only at \(R_G=4\,h^{-1}{\rm Mpc}\). Their genus curves are shown in Figure~\ref{fig:genus-G1}, together with the other results, and the corresponding genus-related statistics are summarized in Table~3. The uncertainties in the TNG100 predictions due to the cosmic variance are estimated from 27 subvolumes extracted from the larger TNG300 simulation, while those for EAGLE are obtained by scaling the TNG100 uncertainties by the square root of the simulation-volume ratio.

Marked discrepancies are found in both simulations for the cluster abundance parameter \(A_C\). The TNG100 and EAGLE values of \(A_C\) exceed the SDSS value by \(4.5\sigma\) and \(7.0\sigma\), respectively. In addition, the EAGLE void abundance parameter \(A_V\) is larger than the SDSS value by \(3.5\sigma\). These discrepancies do not necessarily imply that the subgrid prescriptions adopted in TNG and EAGLE are inconsistent with observations, once the large cosmic variance of the theoretical predictions is taken into account (see the uncertainties in parentheses in Table~3). However, they indicate that the particular TNG100 and EAGLE realizations considered here are statistically inconsistent with the observations, in the sense that they contain too many isolated high-density regions.

In the case of EAGLE, the discrepancy in \(A_C\) may be related to its galaxy--halo connection. EAGLE has been found to preferentially populate galaxies in more massive halos \citep{2017MNRAS.470.1771A} and also produces a relatively high fraction of red galaxies. These related effects can enhance galaxy concentrations in dense environments and may therefore contribute to the elevated \(A_C\).

\section{DISCUSSION AND CONCLUSIONS}
\label{sec:conclusion}
We have measured the three-dimensional genus statistics of nine SDSS DR7 volume-limited galaxy samples over Gaussian smoothing lengths of $R_G=4$--$12~h^{-1}{\rm Mpc}$ and compared them with the HR4 $N$-body simulation and four cosmological hydrodynamical simulations: TNG100-1 and TNG300-1 from IllustrisTNG, FLAMINGO, and EAGLE. Overall, the simulations broadly reproduce the SDSS genus statistics over the smoothing scales where statistically meaningful comparisons can be made. Among them, the HR4 galaxy-assignment model provides somewhat better overall agreement with the SDSS topology than the galaxy populations formed in TNG300 and FLAMINGO.

Compared with the older galaxy-assignment schemes examined by \citet{2010BKAS...35Q..82C}, the modern models considered here show substantially improved agreement with the observed topology. In particular, HR4 agrees well with the SDSS measurements of $\Delta\nu_f$ on small scales, $R_G\leq7~h^{-1}{\rm Mpc}$, whereas the HGC, HOD, and SAM models considered by \citet{2010BKAS...35Q..82C} all showed strong discrepancies at all scales. For $A_C$ and $A_V$, only the HGC model was consistent with the observations in that study, while the HOD model and three implementations of SAM failed to reproduce the measurements. In contrast, all of the models considered here yield $A_C$ and $A_V$ values statistically consistent with the SDSS results.

Despite this overall agreement, the simulations show mild tension with the observed scale dependence of the genus shape parameters. At $R_G\geq8~h^{-1}{\rm Mpc}$, the simulations predict a shallower scale dependence of $\Delta\nu_f$ than observed in SDSS, with somewhat larger deviations in the hydrodynamical simulations than in HR4. The observed $A_V$ also shows a steeper scale dependence than the simulations, although this difference is not individually significant.

The hydrodynamical simulations, TNG300 and FLAMINGO, show somewhat greater departures from the SDSS topology than HR4, although the differences remain modest. The most notable cases are the elevated $A_C$ in TNG300 at $R_G=5
~h^{-1}{\rm Mpc}$ and the more positive $\Delta\nu_f$ in FLAMINGO at $R_G=9~h^{-1}{\rm Mpc}$, differing from the SDSS measurements by about $2.8\sigma$ and $3.7\sigma$, respectively. These results suggest that, even in relatively large-volume hydrodynamical simulations, residual differences in the galaxy topology may persist and provide useful constraints on galaxy-formation modeling.

The topology of the galaxy distributions in the TNG100 and EAGLE realizations shows significant discrepancies from SDSS in the cluster abundance parameter $A_C$, at the $4.5\sigma$ and $7.0\sigma$ levels, respectively, when measured relative to the observational uncertainties. Both simulations contain an excess of isolated high-density regions compared with the observations. When the differences are evaluated relative to the expected cosmic variance of the simulations, estimated from TNG300 subvolumes and scaled to the corresponding volumes, the TNG100 and EAGLE realizations lie $1.7\sigma$ and $2.3\sigma$ from the SDSS value, respectively. Thus, these discrepancies do not necessarily imply a failure of the galaxy-formation prescriptions, but indicate that the particular TNG100 and EAGLE realizations represent uncommon realizations whose galaxy topology differs markedly from that observed in SDSS. This result cautions against treating individual small-volume simulations as representative cosmological volumes without accounting for cosmic variance.

Our Poisson-sampling tests show that finite sampling can affect the measured genus statistics. Survey geometry (Appendix~\ref{sec:validade}) and finite sampling (Appendix~\ref{sec:noise})  therefore must be treated consistently in both the observations and simulations. With its higher galaxy number density, the DESI survey will enable more robust measurements of genus curves, providing a valuable opportunity to test galaxy-formation models.

\section*{Acknowledgments}

This work is supported by the National Natural Science Foundation of China, grant No.12303003, and Yunnan Key Laboratory of Survey Science, grant No.202449CE340002.
CBP is supported by the KIAS Individual Grant PG016903 at the Korea Institute for Advanced Study (KIAS). CBP is also supported by the National Research Foundation of Korea (NRF) grant funded by the Korean government (MSIT; RS-2024-00360385).

\appendix 
\section{Genus Measurements and Validation}
\label{sec:validade}
\subsection{Validation with HR4}
\label{sec:validade1.1}
The SDSS angular mask and finite radial boundaries can modify the smoothed density field near survey boundaries. We account for spatially varying
completeness using the smoothed mask, $\hat{n}^S_g=n^S_g/M^S$, where $n^S_g$ and $M^S$ are the smoothed galaxy density and mask fields, respectively, and retain only pixels where the residual mask effect is sufficiently small. Unlike the approach of \cite{2010BKAS...35Q..82C}, which applies simulation-derived corrections for both the survey mask and RSD to obtain real-space genus statistics, our fiducial analysis excludes strongly affected regions and measures the topology directly in redshift space. This provides a more direct comparison between the observed galaxy distribution and simulations. We therefore analyze the simulations in redshift space as well, avoiding the need for simulation-dependent corrections that may not fully represent the bias and velocity fields of the observed galaxy population.

\begin{figure}
    \centering
     \subfigure{
     \includegraphics[width=0.38\linewidth, clip]{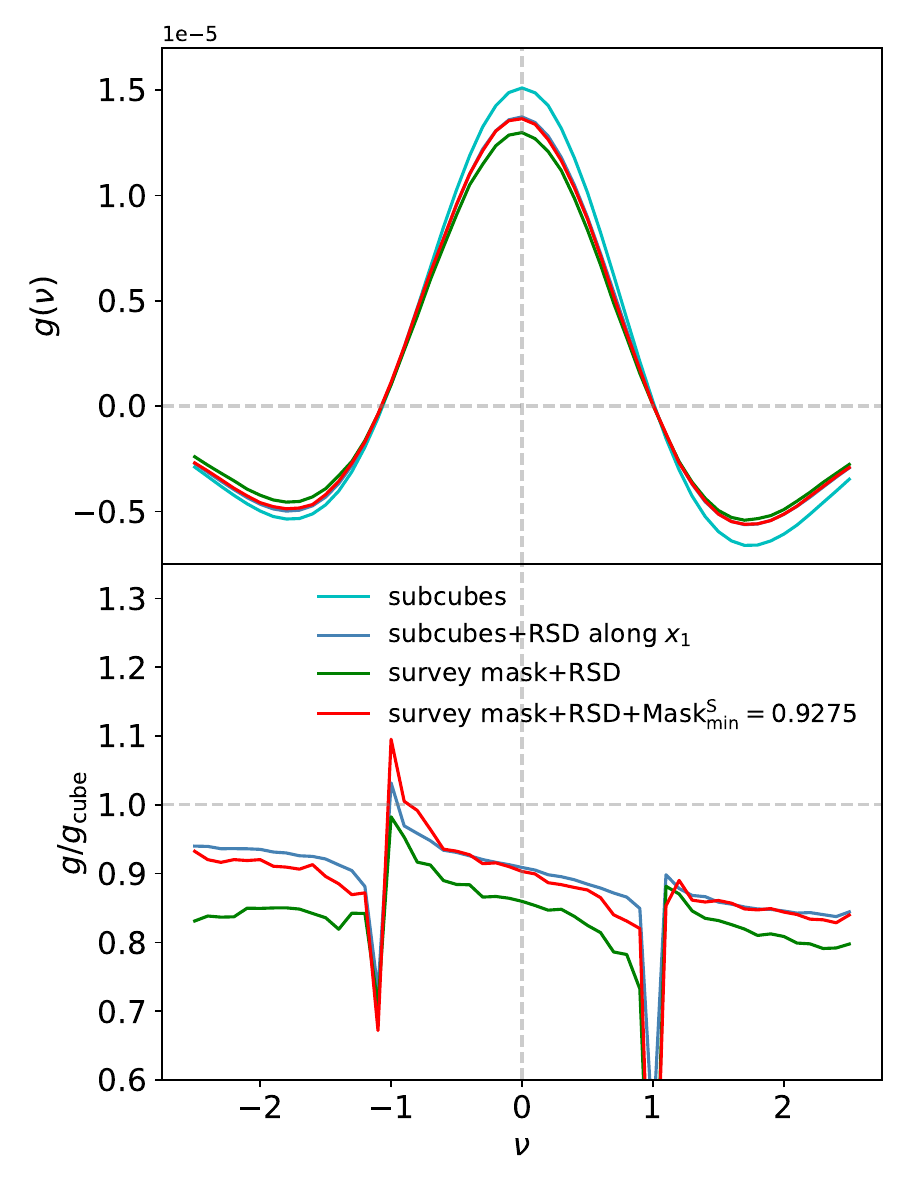}}
     \subfigure{
     \includegraphics[width=0.38\linewidth, clip]{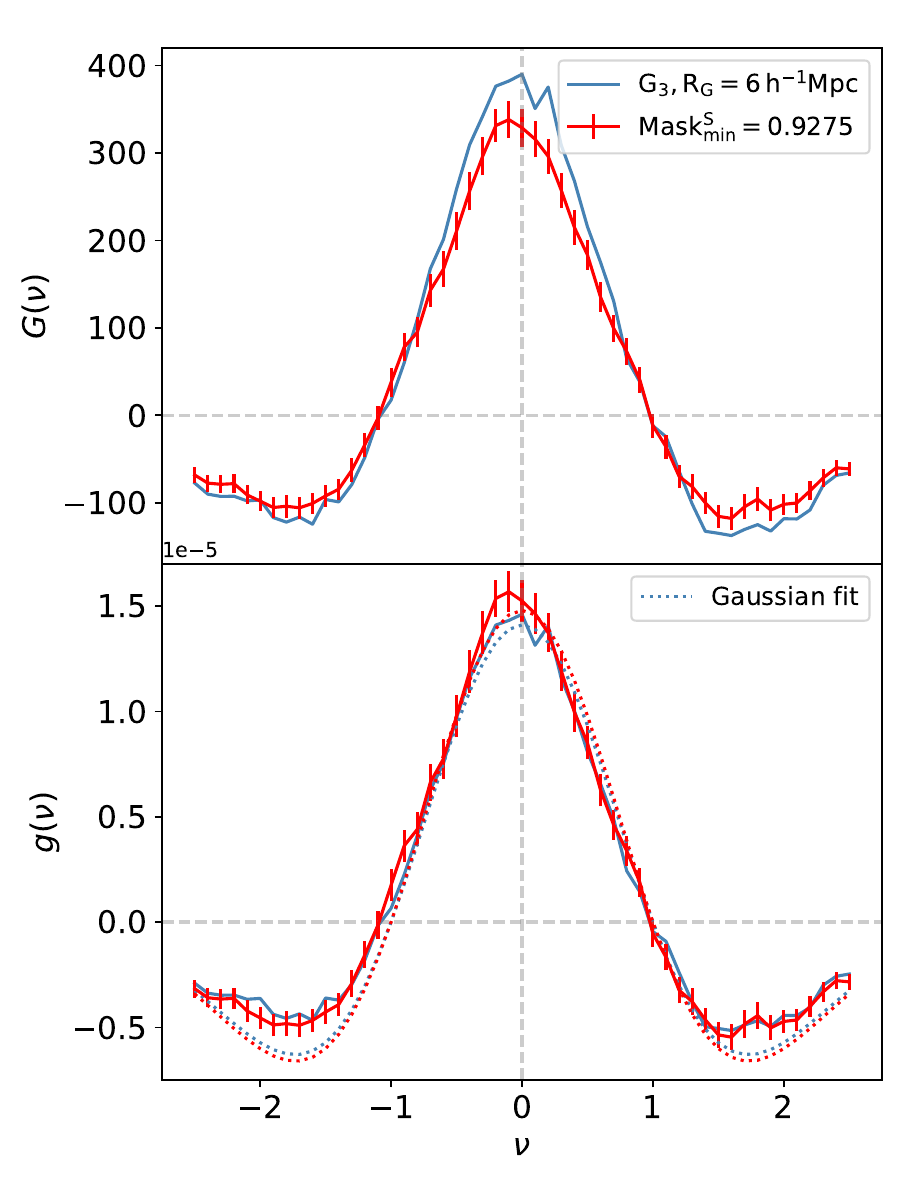}}
   \caption{Validation of the method for mitigating mask \& boundary effects for genus measurements. The left panels are the tests with HR4 mocks, and right panels are from SDSS. The upper left panel shows the HR4 genus curves, and  the lower panel displays the genus curves normalized by the non-RSD subcube measurements, i.e., the cyan curve. These tests are performed on the G3 mock sample with $R_G=6h^{-1}{\rm Mpc}$. The HR4 box has been divided into 125 subcubes, and the curves are obtained by averaging the 125 independent measurements. The cyan and blue curves show the genus measurements without and with the RSD effect; in the latter case, RSD is applied along the $x_1$ direction. The green and red curves corresponds to the realistic mocks, with the red curve employing a cut on the 3D smoothed mask to mitigate the mask effect.  The upper right panel shows the direct measurement $G(\nu)$ of SDSS, and the lower right panel show the genus per unit volume $g(\nu)$. The blue line shows the measurement without accounting for the mask effects introduced by the survey boundary. The red line shows the measurement obtained by selecting pixels with smoothed mask values greater than 0.9275} 
    \label{fig:g6-hr4}
\end{figure}

We validate this procedure using HR4 mock surveys. We divide HR4 into 125 subcubes of side length $640~h^{-1}{\rm Mpc}$ and pad each by $40~h^{-1}{\rm Mpc}$ on each side before smoothing, giving a working volume of $720~h^{-1}{\rm Mpc}$. Pixels outside the original volume are assigned a very negative density after smoothing so that the genus is measured only within the original region.
We use the G3 sample to validate our adopted  measurement procedure. The left two panels of Fig.\ref{fig:g6-hr4} compares four configurations for G3: (i) raw HR4 subcubes in real space, (ii) the same subcubes with RSD along the $x$ direction, (iii) redshift-space mocks with the SDSS angular and radial mask, and (iv) masked mocks with the additional boundary-selection criterion $M^S_{\rm min}=0.9275$. For the second configuration, the redshift-space coordinate is
\begin{equation}
\Tilde{x}=x+v_{x}^{\rm pec}\frac{1+z_{01}}{H(z_{01})},
\end{equation}
where the line of sight is along the $x_1$ direction and $z_{01}$ corresponds to $z=0.1$.

RSD reduces the asymmetry of the reference genus curve, particularly on the high-density side, consistent with the tendency of peculiar velocities to connect neighboring structures. The SDSS mask further suppresses the genus curve, with a stronger effect toward the void side. We therefore retain only pixels with $M^S>0.9275$, corresponding approximately to regions at least $1.5R_G$ from the survey boundary. This selection recovers the reference genus curve well, supporting the use of the same estimator for SDSS and simulation mocks without an empirical correction to the measured genus parameters.

\subsection{Test with observational data}
The right panels of Fig.\ref{fig:g6-hr4} show the genus curve of the DR7 G3 sample at $R_G=6~h^{-1}{\rm Mpc}$ (blue solid line) and its best-fit Gaussian curve (blue dashed line). The observed curve deviates from the Gaussian prediction over the full range considered here, $-2.5\leq\nu\leq2.5$, with a shallower shape at deeper thresholds, indicating a deficit of superstructures relative to a Gaussian random field. This suppression is attributed to the fact that superstructures tend to be more connected and thus exhibit larger characteristic sizes than those in a Gaussian field. We then adopt the methodology described in \S\ref{sec:validade1.1} to mitigate mask effects.  The mask selection reduces the amplitude of $G(\nu)$ because of the smaller effective survey volume, but increases the derived genus density amplitude $g$, indicating that our masking strategy improves the reliability of the measurement and better recovers the intrinsic topological signal.

\section{Poisson sampling effect}
\label{sec:noise}
\begin{figure*}
    \centering
     \subfigure{
     \includegraphics[width=0.7\linewidth, clip]{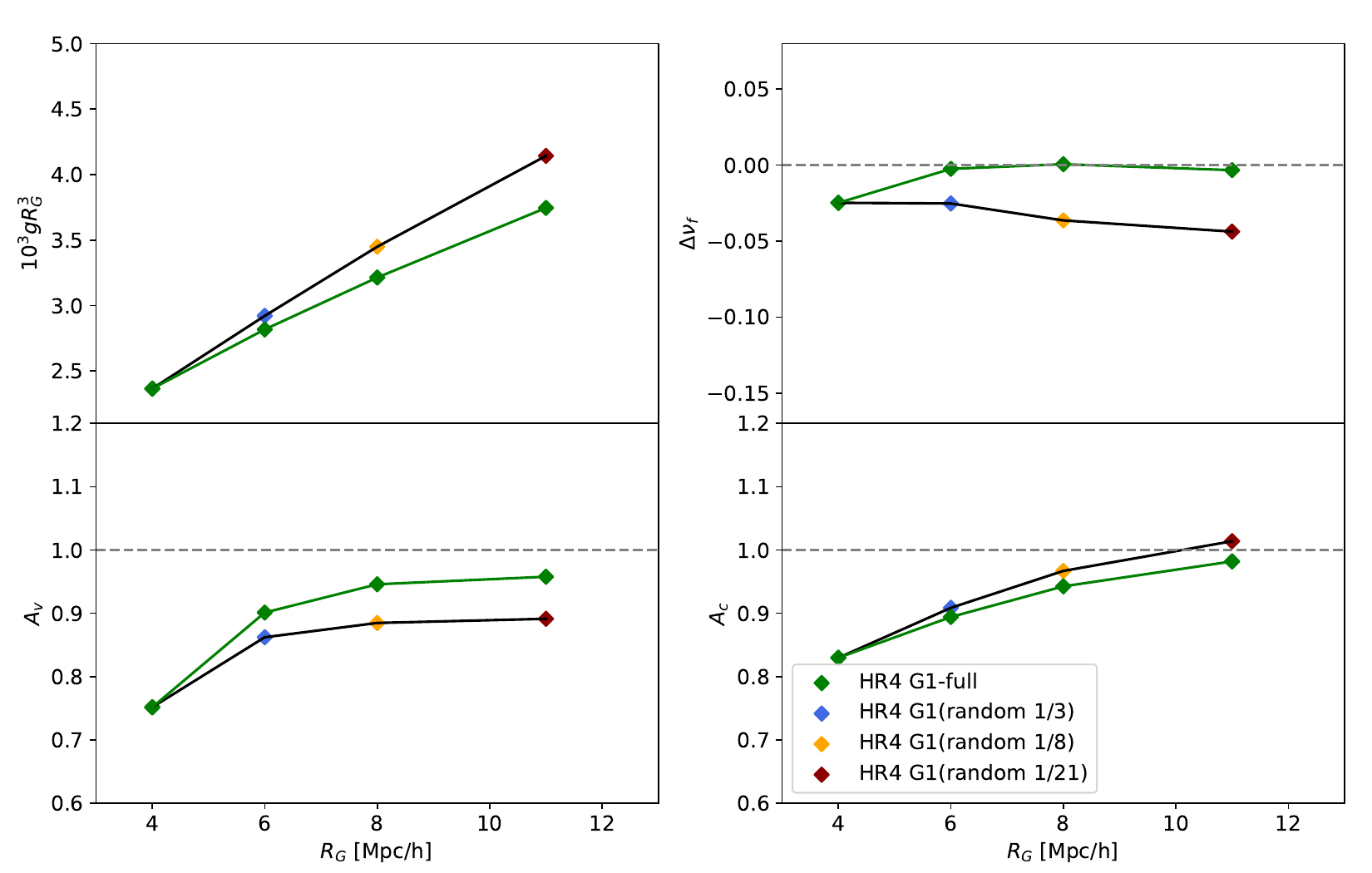}}
   \caption{Similar to Fig.\ref{fig:shape}, but showing the effects of finite sampling on the genus-related parameters, illustrated using the HR4 G1 sample. Green diamonds connected by green lines denote results adopting the full G1 sample galaxy number density, while colored diamonds connected by black lines show randomly subsampled catalogs with number densities reduced to $\frac{1}{3}$,$\frac{1}{8}$ and $\frac{1}{21}$, respectively. The systematic offsets between the black and green lines arise from shot noise.} 
    \label{fig:short-noise}
\end{figure*}

The volume-limited samples in Fig.~\ref{fig:shape} exhibit a systematic luminosity dependence in their genus statistics. However, this apparent dependence may arise from multiple effects. For example, brighter samples are increasingly sparse and thus more susceptible to Poisson sampling, while galaxy bias varies with luminosity and may also be scale-dependent. Together, these effects can modify the shape of the galaxy power spectrum\footnote{The observed galaxy power spectrum can be expressed as $P_{\rm obs}=b^2P_{\rm mm}+P_{\rm shot}$, where $P_{\rm mm}$ is the matter power spectrum, $b$ is the galaxy bias, and $P_{\rm shot}=1/\bar n_g$ is the Poisson shot-noise contribution.},  thereby affecting the measured genus statistics. Here we use the HR4 G1 sample as an example to show the impact of Poisson sampling on the measured genus statistics.

We randomly downsample the full sample to $1/3$, $1/8$, and $1/21$ of its original number density, yielding mean galaxy separations of approximately $\bar{d}\sim4$, $6$, $8$, and $11~h^{-1}{\rm Mpc}$, respectively. This procedure keeps the underlying galaxy population and its bias fixed, while varying only the sampling density. The genus statistics are measured in redshift space, with RSDs applied along the $x$-axis. The results are shown in Fig.~\ref{fig:short-noise}. For the G1-full sample, increasing $R_G$ drives  $\Delta\nu_f\to0$ and $A_V,A_C\to1$ (green solid lines). Relative to these results, downsampling increases $gR_G^3$, makes $\Delta\nu_f$ more negative, lowers $A_V$, and slightly increases $A_C$ (black solid lines). The different responses of these parameters can be understood in terms of the highly inhomogeneous galaxy density field. In underdense regions, where the local galaxy density is low, Poisson sampling has a relatively greater impact. Individual galaxies in these regions can be smoothed into localized overdense features, potentially modifying the void topology. Meanwhile, downsampling may also break up overdense structures into more isolated clusters, further modifying the genus curve. By comparison, $A_C$ is less affected because overdense regions contain more galaxies and are therefore less susceptible to Poisson fluctuations.

Overall, Poisson sampling has a non-negligible impact on the genus measurements of the SDSS DR7 samples. In our analysis, we do not apply an HR4-based Poisson correction to the SDSS measurements, as the galaxy populations in HR4 may differ from those in the observations. Instead, we match the number densities of the mock catalogues to those of the observed samples, ensuring a consistent comparison at comparable sampling densities. Denser surveys such as DESI, with higher galaxy number densities, will help mitigate finite-sampling effects and enable more robust genus measurements.

\bibliography{genus}
\end{document}